\documentclass[aps,twocolumn,showpacs,preprintnumbers,amsmath,amssymb,nofootinbib,prc,10pt]{revtex4-1}

\usepackage{graphicx}

\newcommand{\beq}{\begin{equation}}
\newcommand{\enq}{\end{equation}}
\newcommand{\beqn}{\begin{eqnarray}}
\newcommand{\enqn}{\end{eqnarray}}
\newcommand{\al}{\alpha}
\newcommand{\be}{\beta}
\newcommand{\om}{\omega}
\newcommand{\dg}{\dagger}
\newcommand{\lm}{\lambda}
\newcommand{\f}[2]{\frac{#1}{#2}}
\renewcommand{\d}{\mathrm d}
\newcommand{\nn}{\nonumber}

\begin{document}

\title{Symmetric nuclear matter with chiral three-nucleon forces in the self-consistent Green's functions approach}

\author{Arianna Carbone}
\affiliation
{Departament d'Estructura i Constituents de la Mat\`eria and Institut de 
Ci\`{e}ncies del Cosmos, Universitat de Barcelona, E-08028 Barcelona, Spain}

\author{Arnau Rios}
\affiliation{Department of Physics, Faculty of Engineering and Physical Sciences, University of Surrey, Guildford, Surrey GU2 7XH, United Kingdom}

\author{Artur Polls}
\affiliation
{Departament d'Estructura i Constituents de la Mat\`eria and Institut de 
Ci\`{e}ncies del Cosmos, Universitat de Barcelona, E-08028 Barcelona, Spain}

\date{\today}

\begin{abstract}
We present calculations for symmetric nuclear matter 
using chiral nuclear interactions within the Self-Consistent Green's Functions approach in the
ladder approximation. 
Three-body forces are included via effective one-body and two-body interactions, 
computed from an uncorrelated average over a third particle. 
We discuss the effect of the three-body forces on the total energy, 
computed with an extended Galitskii-Migdal-Koltun sum-rule, as well as on
single-particle properties.
Saturation properties are substantially improved when three-body forces are included, 
but there is still some underlying dependence on the SRG evolution scale. 	
\end{abstract}

\maketitle

\section{Introduction}

The study and understanding of infinite nuclear matter has always been a fundamental goal of nuclear physics \cite{Ring84}.
A major advantage is the simplicity of the system at the one-body level, which allows for tests of 
approximations on the many-body sector. 
The hope is that these approximations allow for a further understanding of correlations in finite systems. 
From an \emph{ab initio} point of view, one would like to start from two-nucleon forces (2NF) which describe accurately
two-body properties, \emph{i.e.} nucleon-nucleon ($NN$) phase shifts and deuteron data. This is generally achieved via fits to 
the Nijmegen database \cite{Sto93}. 

Unfortunately, whatever realistic 2NF is used in a many-body calculation, the saturation properties of nuclear matter fail to be reproduced \cite{Muther2000,Dewulf2003}. Saturation densities are too large and saturation energies too attractive, with calculations falling in the so-called Coester band. It is generally expected that the inclusion of three-nucleon forces (3NF) will help cure this deficiency. This bears some similarity to the case of light nuclei, whose binding energies are not correctly predicted when computed with 2NF only \cite{Wiringa2002}. In the few-body case, however, the theory based on 2NF only underbinds experimental data, whereas nuclear matter is generally overbound. Consequently, 3NFs have been mostly devised to provide further attraction in light systems (or small densities) and repulsion for the infinite system (high densities) \cite{Piep01}. The attractive part is often modeled on the seminal two-pion exchange force of Fujita and Miyazawa \cite{Fuj57}.

First principles calculations should be based on a consistent hamiltonian, with 2NF and 3NF that include the same ingredients and are computed within the same model space. This has not been necessarily the case in the past, where forces, particularly at the three-body level, had phenomenological ingredients \cite{Piep01}.
A way out of \emph{ad hoc} parametrizations has been provided by effective field theories (EFT) applied to low-energy quantum chromodynamics. 
This approach exploits the separation of momentum scales at low energies to define nucleons and pions as the only active degrees of freedom. Their interactions are included via a power counting originally proposed by Weinberg \cite{Wein90,*Wein91} and further developed later on \cite{Ord94,*Ord96}. We note here that Weinberg power counting is not necessarily the only way to tackle these issues \cite{Kapl96,*Kapl98b,*Kapl98,Nog05,Epel09,Epel09b}. 
In the end, chiral EFT ($\chi$EFT) provides a consistent set of hamiltonians at the two-, three- and four-body level which can be systematically improved by including higher orders in an expansion of momenta over a large chiral scale.

Exploiting chiral nuclear forces, we present in this paper results for infinite nuclear matter considering both 2NF and 3NF. We perform calculations using the Entem-Machleidt 2NF \cite{Ent03}, a chiral two-body force based on terms up to fourth order (N3LO) in $\chi$EFT within Weinberg's power counting \cite{Mac11}. We also employ a 3NF obtained at third order (N2LO), which is reduced to a density-dependent effective interaction following the prescription introduced by Holt and collaborators \cite{Holt10}. This represents in some sense the lowest-order contribution of the three-body force in the many-body system. Note that, compared to the 2NF, the N2LO 3NF depends on two additional low-energy constants which need to be obtained in the few-body sector \cite{Nogga2006,Nav07,Marcucci2013}. Calculations will therefore depend on these two constants that change with the renormalization scale.  

In the past, most calculations for infinite matter with 3NF have been performed using an averaged, density-dependent 2NF that arises from three-body physics. This is then added to the original 2NF to provide an interaction that effectively includes both two- and three-body effects. 
Within the Correlated Basis Function theory, phenomenological density-dependent 3NF have been used for over 30 years \cite{Lag81,Fri81,Fan84,*Fan87,Fab89,Ben89,*Ben92}. 
Recently, further static and dynamical correlations have been introduced in the density-dependent force within the Fermi-Hypernetted-Chain approach \cite{Lov11,*Lov12}, both for Urbana IX \cite{Piep01} and for a local version of the N2LO chiral 3NF \cite{Nav07}. 
Urbana IX has also been used extensively within the Brueckner--Hartree--Fock (BHF) approach \cite{Bal99,Zuo02,*Zuo02b,Vid09}. Within this approach, most results have been derived from a density-dependent 2NF following the prescription of Ref.~\cite{Gran89}. Modern calculations also include a consistent defect function in the averaging procedure for other 3NF \cite{Li08,*Li08b}. 

Our calculations are performed within the self-consistent Green's functions (SCGF) framework \cite{Dick04}. This is a general diagrammatic approach, based on the perturbative expansion of the single-particle (SP) propagator \cite{Dick05}. The non-perturbative nature of the interaction is taken into account via an in-medium ladder resummation. The inclusion of 3NF within this framework requires a series of formal developments. These will be dealt with in a forthcoming publication \cite{Car}. For practical purposes, and as a first consistency test, we will include the 3NF in the simplest possible way. This involves a SP average over the third particle, which we will borrow from Holt \emph{et al} \cite{Holt10}. Note, however, that we go one step further with respect to previous SCGF calculations, since we take into account the differences between the 3NF contribution at the lowest order (Hartree-Fock level) and at all successive orders (dispersive contributions).

Our results are relevant in view of recent advances in \emph{ab initio} calculations of finite nuclei based on the very same chiral interactions \cite{Hagen2012,Ro12,Soma2013}. This new generation of predictions rely on a similarity renormalization group (SRG) transformation to make the 2NF more tractable in a many-body framework \cite{Bog07,Bog10}. The SRG evolution, however, introduces induced 3NF and these need to be consistently taken into account. 
Just recently, satisfactory results for the binding energies of isotopes in the nitrogen, oxygen and fluorine chains have been obtained using this 3NF-expanded SCGF formalism \cite{Cip13}, where a SRG-evolved chiral hamiltonian, including both induced and pre-existing 3NF, has been used. 

An important consistency check at this level should be that of the saturation properties of symmetric nuclear matter (SNM). If these are well reproduced with the same underlying interaction, one expects the radii and binding energies of nuclei to come out close to experiment \cite{Muther2000}. 
A full SRG evolution of 2NF and 3NF has been achieved only recently in momentum space, which is the relevant basis for nuclear matter calculations \cite{Hebeler2012,Heb13,*Wendt2013}. Previous applications employed a simpler strategy, whereby the SRG evolution was performed on the 2NF alone \cite{Hebeler2010,Heb11} and the low-energy constants (LEC) of the 3NF were refitted. At each resolution scale, the fit is performed to reproduce the properties of the three- and four-body system. The underlying hypothesis is that the effect of induced 3NF is fully accounted for in the renormalization of these LECs. In other words, one assumes that the functional dependence of the original and the induced 3NF are the same and the SRG evolution only affects the pre-factors. 

To test whether this is the case or not, one should perform calculations with the original Hamiltonian and with evolved nuclear forces. Scale invariance is to be expected whenever the many-body truncations are under control. Alternatively, the breaking of the invariance under the SRG evolution scale can be taken as a measure of the theoretical error, arising from either missing many-body correlations or numerical limitations \cite{Bog10}. In general, however, invariance can only be tested by performing calculations in a wide range of resolution scales. Unfortunately, if one wants to stay within the perturbative regime, the range of evolution scales that can be used is somewhat limited, since the original hamiltonian, at least in the two-body sector, is non-perturbative. In this context, it seems important to develop and implement a many-body scheme that can treat 2NF and 3NF within a  non-perturbative framework. An attempt to this issue has been presented recently \cite{Oll10,*Lac11}, in which a non-perturbative approach to the nucleon and pion self-energy was performed up to next-to-leading order in $\chi$EFT for applications in infinite nuclear matter.

In this paper, we shall focus on the effect of 3NF on bulk and single-particle properties in symmetric nuclear matter. We will compare our results with BHF calculations based on the same underlying interactions. 
In Section II, we discuss briefly the modifications implemented in the SCGF approach when 3NF are considered. 
Results for the bulk and single-particle properties of nuclear matter are presented in Section III. 
Summary and conclusions are presented in Section IV.   

\section{Self-consistent Green's functions with three nucleon forces}

In nuclear matter without further renormalization, the short-range of the interaction has to be treated non-perturbatively. Within Green's functions, the ladder re-summation has been traditionally employed to take into account the strong short-distance repulsion \cite{Ramos1989,Frick2003,Rios2008,Soma08}. This amounts to an infinite sum of diagrams, which gives access to an effective in-medium interaction, the so-called $T-$matrix. The method goes beyond traditional BHF calculations by extending the sum to include not only particle-particle, but also hole-hole intermediate states. Moreover, particles are dressed at all stages by imposing self-consistency at the propagator level, using the Dyson equation. This ensures thermodynamical consistency and, consequently, the fulfillment of the Hugenholtz-van Hove theorem \cite{Hugenholtz1958}. 

So far, most ladder SCGF calculations have been performed using 2NF only \cite{Ramos1989,Frick2003,Rios2008}. The Krakow group has presented results using the Urbana IX force, with an average that includes the effect of correlations \cite{Soma08,Soma09,*SomaPhD}. Unfortunately, as we shall see in the following, such calculations are flawed due to a missing factor in diagrams involving Hartree-Fock-like three-body terms.
A more thorough analysis on how the SCGF method is extended to account for 3NF will be presented in the future \cite{Car}. There, we shall also comment upon the fact that the Hartree-Fock and the dispersive contributions of the 3NF to the self-energy actually require different averaging procedures. 

Density dependent effective 2NF have been recently constructed from chiral 3NF at N2LO in Refs.~\cite{Holt10,Heb10}. In both cases, the average is performed using a non-interacting propagator. This yields semi-analytical expressions for the effective 2NF matrix elements, which can then be used in perturbative calculations of infinite \cite{Heb11} and finite nuclear systems \cite{Holt09}. Note that the differences between the two density-dependent forces are small and due mostly to different assumptions in the averaging procedure. In contrast, a recent BHF calculation based on the same chiral 3NF  but with an alternative averaging procedure, indicates a very strong overbinding of nuclear matter \cite{Li12}. The results presented here do not appear to support such claims. 

\begin{figure}
\begin{center}
\includegraphics[width=0.7\linewidth]{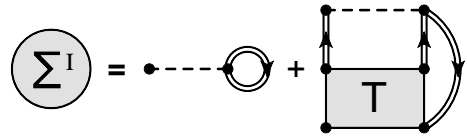}
\caption{The ladder irreducible self-energy can be obtained from the sum of a generalized Hartree-Fock term and a dispersive contribution. Dashed lines represent two-nucleon interactions and double lines with arrows are self-consistently dressed single-particle propagators.}
\label{fig:self_en_2b}
\end{center}
\end{figure}

The SCGF method aims at calculating a dressed expression for the SP propagator in the medium, $G_{\al\al'}(\om)$, where $\al$ represents the quantum number of a SP excitation and $\om$, its energy. The SP Green's functions fulfills the Dyson equation, 
\beq
\label{Dyson}
G_{\al\al'}(\om)=G^{(0)}_{\al\al'}(\om)+\sum_{\be\be'}G^{(0)}_{\al\be}(\om)\Sigma^I_{\be\be'}(\om)G_{\be'\al'}(\om)\,,
\enq
where $G^{(0)}$ corresponds to an undressed, in-medium SP propagator. Eq.~(\ref{Dyson}) is solved iteratively until self-consistency is achieved. The irreducible self-energy operator,  $\Sigma^I(\om)$, 
contains one-particle irreducible diagrams. In the ladder approximation and when considering only 2NF, the irreducible self-energy is obtained from two distinct terms. The first represents a generalized Hartree-Fock contribution, where the bare interaction is integrated once over a dressed propagator.  The second contribution depends on energy and it is often called dispersive term. It conceals all the higher-order contributions in an integration of the in-medium ladder $T$-matrix over a complete propagator. 
A diagrammatic representation of these two terms is presented in Fig.~\ref{fig:self_en_2b}.
If the calculation is performed including 3NF, one has to evaluate correctly the diagrammatic expansion for the self-energy \cite{Car}. The traditional ladder approximation can be extended to include the lowest-order effects associated to 3NF. In second quantization, a Hamiltonian including a one-body (kinetic), two-body and three-body interaction reads:
\beqn
\label{H}
H &=& \sum_{\mu} T_\mu\, a^\dg_\mu a_\mu
+
\f 1 4 \sum_{\substack{\mu\nu\\ \bar \mu \bar \nu}}V_{\mu\nu \bar \mu \bar \nu}\, a_\mu^\dg a_\nu^\dg a_{ \bar \nu} a_{ \bar \mu}
\\\nn &+& 
\frac{1}{36}\sum_{\substack{\mu\nu\lm \\  \bar \mu \bar \nu \bar \lm}} W_{\mu\nu\lm \bar \mu \bar \nu \bar \lm}\,
a_\mu^\dg a_\nu^\dg a_\lm^\dg a_{ \bar \lm} a_{ \bar \nu} a_{ \bar \mu}\,,
\enqn
where a basis set which diagonalizes the kinetic operator has been used. $a^\dg_\mu$ and $a_\mu$ are
the operators which respectively create or destroy a particle in state $\mu$. $T_{\mu}$ correspond to
the matrix elements of the kinetic operator, whereas $V_{\mu\nu \bar \mu \bar \nu}$ and $W_{\mu\nu\lm \bar \mu \bar \nu \bar \lm}$ are antisymmetrized elements of the 2NF and 3NF. We note here that, for all practical purposes it is enough to anti-symmetrize the ket state. 

The diagrammatic expansion associated to both the two-body and the three-body interaction can be obtained from the usual perturbartive expansion of the single-particle propagator \cite{Dick05}. 
The number of diagrams is substantially larger than that obtained in the case where only 2NF are considered.
This does have implications in both perturbative and non-perturbative calculations. For the former, many more diagrams would need to be considered at each order in perturbation theory. Note that these would include both pure 3NF terms as well as mixed terms, including combinations of 2NF and 3NF. For non-perturbative calculations, one would like to include 3NF in the resummation procedure somehow. Both issues can be tackled effectively by a generalization of the technique of normal ordering of the Hamiltonian \cite{Ots10,Ro12}. 

\begin{figure}
\begin{center}
\includegraphics[width=0.9\linewidth]{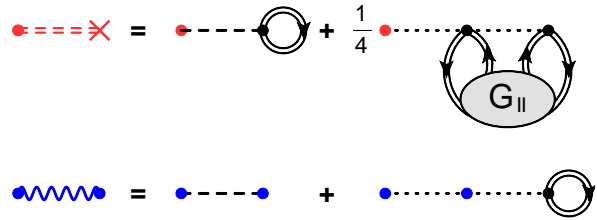}
\caption{(Color online) Diagrammatic definition of the one-body (double-dashed line) and two-body (wiggly line) effective interactions. The one-body contribution is the sum of the generalized Hartree-Fock term of the 2NF plus a term containing the two-body propagator, $G_{II}$; dotted line represents three-nucleon interactions. The $1/4$ factor counts the equivalent lines in the latter diagrams. The two-body effective force is the sum of the 2NF and a 3NF averaged over one particle. }
\label{fig:effective}
\end{center}
\end{figure}

\begin{figure}
\begin{center}
\includegraphics[width=0.7\linewidth]{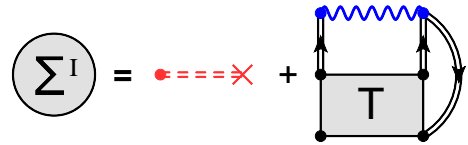}
\caption{(Color online) Ladder irreducible self-energy employed in this work, using the effective interactions defined in Fig.~\ref{fig:effective}.}
\label{fig:self_en_3b}
\end{center}
\end{figure}

\begin{figure}
\begin{center}
\includegraphics[width=0.7\linewidth]{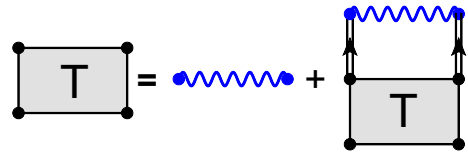}
\caption{(Color online) T-matrix employed in this work, using the effective interactions defined in Fig.~\ref{fig:effective}.}
\label{fig:tmat_3b}
\end{center}
\end{figure}

In this approach, one defines effective one- and two-body interactions by averaging over two- or one-particle states with fully dressed propagators, as depicted in Fig.~\ref{fig:effective}. These incorporate effectively an infinite sub-series of terms of the original perturbative expansion of the propagator.  The advantage of using effective interactions lies in the spontaneous grouping of diagrams, which facilitates the enumeration of contributions in the self-energy. When defining such effective interactions, however, one can devise a new diagrammatic expansion which has fewer terms than the original one.  To be more specific, one needs to consider only diagrams which are interaction irreducible \cite{Car,Cip13}.  The reorganization leads to the same diagrammatic content as the original expansion if all diagrams are considered.
In particular, we will take into account the extension of the ladder resummation technique depicted in Fig.~\ref{fig:self_en_3b}. This includes both generalized Hartree-Fock terms associated to the 2NF and 3NF, as well as non-perturbative contributions arising from the ladder resummation of the two-body effective interaction. 
We show in  Fig.~\ref{fig:tmat_3b} the $T$-matrix that appears in the dispersive term of Fig.~\ref{fig:self_en_3b}. Note that this is calculated using effective two-body interactions only. For simplicity, we neglect all contributions coming from pure 3NF interaction irreducible diagrams. A first attempt to quantify the importance of these has been made in Ref.~\cite{Kaiser2012}.

It must be pointed out that in the present work we use a one-body effective interaction that is an approximation to that depicted in Fig.~\ref{fig:effective}. As shown in Fig.~\ref{fig:effective}, the complete expression for the second term arises from the contraction of the three-body term with a correlated pair state and is beyond the scope of this first exploratory work \cite{Car}. Instead, in the present calculations, the matrix elements for the second term of the one-body effective interaction are given by the lowest-order term in the expansion of the two-body propagator, i.e. the Hartree-Fock approximation for the $G_{II}$:
\beqn
\label{ueff}
\tilde{U}_{\mu\bar \mu} &=& \sum_{\nu \bar \nu} \left[  
V_{\mu \nu \bar\mu \bar \nu} 
+ \frac{1}{2} \sum_{\lm \bar \lm} W_{\mu\nu\lm \bar\mu \bar \nu \bar \lm} 
\rho_{\lm \bar \lm} \right]  \rho_{\nu \bar \nu} \, .
\enqn
Note that, within this approximation, the pre-factor in front of the three-body force is larger by a factor of $2$ with respect to that of Fig.~\ref{fig:effective}. This is a consequence of the equivalence of the direct and exchange terms in this lowest-order approximation.
In the previous expansion, the one-body density is given by the SP propagator,
\beqn
\rho_{\nu \bar \nu} =  i\hbar G_{\nu \bar \nu}(t-t^+) = \int_{C \uparrow} \d\om \, G_{\nu \bar \nu}(\om) \, .
\enqn
For an infinite system, where momentum is a good quantum number, the one-body density reduces to the momentum distribution, $\rho_{k k'} =  \delta_{k,k'} n(k)$. The two-body effective interaction is given by the sum of the original 2NF and the density-dependent term arising from the original 3NF:
\beqn
\label{veff}
\tilde{V}_{\mu\nu \bar \mu \bar \nu} &=&  V_{\mu\nu \bar \mu \bar \nu} 
 + \sum_{\lm \bar \lm}W_{\mu\nu\lm \bar \mu \bar \nu \bar \lm} \rho_{\lm \bar \lm} \, .
\enqn
Note that the previous equations hold for anti-symmetrized matrix elements. In this scheme, the three-body force enters the calculation of both effective interactions in a similar way, via a one-body average. Hence, the only input that is actually needed is the average of $W$ over the density, which we take from Ref.~\cite{Holt10}.

We would like to stress here that the pre-factor accompanying the three-body force in Eqs.~(\ref{ueff}) and (\ref{veff}) is different. The one-half factor arises from the presence of two equivalent lines, starting and ending in the same vertex of the three-body force in the two-body average. In practice, this means that the density-dependent matrix elements are summed differently in the Hartree-Fock-like term and the dispersive contribution to the self-energy \cite{Heb10}. As pointed out in Ref.~\cite{Heb10}, this effect has not been properly taken into account in previous calculations of infinite matter using either the BHF or the SCGF methods \cite{Bal99,Zuo02,Soma08,Li12}. 

In the following, we will use the density-dependent 3NF of Holt and collaborators \cite{Holt10} as a first approximation to the one-body average. Let us mention here that the average is performed over an uncorrelated density and hence one should replace $\rho \to \rho_0$ in the terms containing $W$ in Eqs.~(\ref{ueff}) and (\ref{veff}). Based on the typical structure of the momentum distribution in nuclear matter, we expect that correlations will only bring small corrections. Our preliminary analysis indicates that the terms proportional to the density in the effective interaction do not change when correlations are taken into account \cite{Holt10}. We expect that the remaining contributions will be modified within $10-20 \%$ by the correlation effect associated to the momentum distribution. Additional, but small corrections, should arise from the inclusion of the full two-body propagator in the one-body effective interaction. Thus, the influence of 3NF effects will be qualitatively captured in the calculations presented below. We are presently working on our own averaged density-dependent 2NF with correlated SP propagators.

We have exploited the similarities between the schemes depicted in Fig.~\ref{fig:self_en_2b} and Fig.~\ref{fig:self_en_3b} to extend our existing SCGF numerical routines to include 3NF \cite{Frick2003,Rios2008}. Calculations are performed at finite temperature, to avoid pairing instabilities which arise below a given critical temperature \cite{Ramos1989,Alm1996}. When a converged solution is reached for the self-energy, one can obtain the SP spectral function, $A(k,\om)$. This gives access to all the SP properties of the system and, in particular, to the momentum distribution,
\beq
\label{mom}
n(k)=\int\frac{\d\om}{2\pi} A(k,\om)f(\om)\,,
\enq
where $f(\om)$ is the Fermi-Dirac distribution function. 
We will concentrate on the low-temperature regime, so we do not expect to find large corrections due to thermal effects. When necessary, we will minimize thermal effects by proper extrapolations to zero temperature. Note, however, that the uncorrelated propagator in the 3NF average, $\rho^0$, is computed at zero temperature. 

When the hamiltonian is formed of 2NF only, one can compute the total energy of the system from the single-particle spectral function using the Galitskii-Migdal-Koltun (GMK) sumrule \cite{Dick05}. With 3NF, one can obtain the total energy of the system using the very same ideas:
\beq
\label{sumrule}
\frac{E}{A}=\frac{\nu}{\rho}\int\frac{\d^3 k}{(2\pi)^3}\int\frac{\d\om}{2\pi}\f 1 2 \left\{\frac{k^2}{2m}+
\om\right\}A(k,\om)f(\om) - \f 1 2 \langle W\rangle\,.
\enq
Here, we have chosen to separate the GMK sum-rule into a contribution containing the SP spectral function plus a pure three-body term. This is the expectation value of the three-body interaction, $\langle W\rangle$, over the full many-body wave function. In our calculations, we estimate this contribution from the three-body term in Eq.~(\ref{ueff}) and integrate it over a correlated momentum distribution \cite{Cip13}, taking into account an additional 1/3 factor due to symmetry properties of  $\langle W\rangle$. Alternatively to Eq.~(\ref{sumrule}), one could have employed an expression using the two-body propagator, $G_{II}$, without any explicit reference to the three-body force \cite{Car}. 

To compare with previous literature as well as to provide a cross-check to our results, we will also present a series of BHF results which have been obtained from an independent code. The inclusion of the 3NF in this approach is somewhat different to what has been done traditionally. Since the dispersive contribution is not usually computed in a $G$-matrix context, we cannot easily separate the one- and the two-body effective interactions to take into account the different pre-factors on the averaged force. Instead, we perform a complete BHF calculation without the $1/2$ factor in front of the three-body part of Eq.~(\ref{ueff}) and then introduce the proper correction to recover the Hartree-Fock contribution to the total energy per nucleon.

\section{Results}

\subsection{Bulk properties}

In the present work, we include 3NF in the SCGF formalism as a density-dependent 2NF following the prescription by Holt \emph{et al.} \cite{Holt10}. This 2NF is derived from the 3NF contributions appearing at N2LO in $\chi$EFT, namely two-pion exchange, iterated one-pion and contact terms. The LECs appearing in the two-pion exchange term are the same as those appearing in the original 2NF at N3LO. These are fixed by experimental $NN$ phase-shifts and deuteron properties \cite{Ent02}. In contrast, the two LECs appearing in the one-pion and contact 3NF terms, $c_D$ and $c_E$, remain undetermined and have to be fit to further experimental values in the few-body sector. The values we use, $c_D=-1.11$ and $c_E=-0.66$, are obtained from ground-state properties of $^3$H and $^4$He with the same, unevolved 2NF and 3NF \cite{Nogga2006}. The cutoff applied on the 3NF is set at $\Lambda_\mathrm{3NF}=500$ MeV, as presented in Refs.~\cite{Nogga2006,Nav07}, and the functional form of the regulator function follows the one given by Holt \emph{et al.} \cite{Holt10}. The determination of these constants from different few-body observables has been discussed in the literature \cite{Nogga2006,Nav07,Marcucci2013}.
We use partial waves up to $J=4$ ($J=8$) in the dispersive (Hartree-Fock) contribution.
The original density-dependent force is given only for diagonal elements, $p=p'$, in relative momentum. To extend these to off-diagonal momenta, we follow the prescription of replacing $p^2 \to \frac{1}{2}(p^2+p'^2)$, as proposed in Ref.~\cite{Holt09,Holt10}.

\begin{figure}
\begin{center}
\includegraphics[scale=0.55]{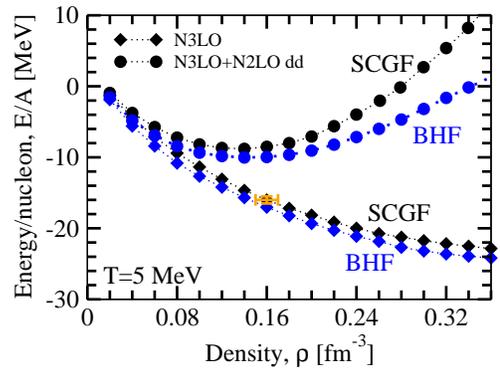}
\caption{(Color online) SCGF (black) and BHF (blue) results for the energy per nucleon of SNM as a function of the density at a temperature of $T=5$ MeV. Diamonds represent calculations including only 2NF, whereas circles include 2NF and 3NF as described in the text. The cross gives the empirical saturation value of nuclear matter. }
\label{ener}
\end{center}
\end{figure}

Fig.~\ref{ener} shows the energy per particle for both the SCGF and the BHF approaches as a function of density for a temperature of $T=5$ MeV. We show in the same figure the results obtained with 2NF only (diamonds) as well as those obtained with 2NF and 3NF (circles). 
For the 2NF-only results, the curves saturate at an unrealistically high density ($\sim 0.42$ fm$^{-3}$, beyond the limit of the Figure) and far too attractive energy ($\sim -22$ MeV). As expected, the SCGF results are more repulsive than the BHF ones due to hole-hole scattering. This difference is however hardly noticeable for the original N3LO potential.

For both approaches, the inclusion of the 3NF is crucial to improve the saturation properties of nuclear matter. We obtain a qualitatively good saturation density of $\rho \sim 0.14$  fm$^{-3}$ and a slightly repulsive saturation energy of $\sim 10$ MeV. The inclusion of 3NF brings in additional differences between BHF and SCGF results at high densities. In particular, at $\rho=0.32$ fm$^{-3}$, SCGF results are $\sim 5$ MeV more repulsive than BHF. The difference in the many-body results can be attributed to two effects. On the one hand, the density-dependent 2NF might bring in additional short-range and tensor correlations with respect to the original force. Hence, the difference in many-body correlations shows up more clearly.  On the other hand, the treatment of the 3NF is slightly different in both cases. Whereas in the SCGF this is included in a consistent way through the complete calculation, the 3NF in the BHF is only introduced consistently at the Hartree-Fock level. 

Let us stress here that our 2NF BHF calculations agree qualitative with those presented in Ref.~\cite{Li12}. However, once the density-dependent force is included, we find a completely different behavior. In our case, the density-dependent 2NF provides repulsion and brings the saturation density to a very reasonable value. In contrast, the averaging procedure given by Li \emph{et al.} suggests a collapsing scenario, where the energy decreases steadily with increasing density even when 3NF are included \cite{Li12}.

Additional sources of theoretical errors are expected to arise at this level of approximation. First, the averaging procedure is performed at the one-body level and over an uncorrelated, zero-temperature Fermi sea. As mentioned earlier, this average can be improved and is expected to produce a small, $10-20 \%$ reduction of three-body effects. We foresee that the two-body average should induce a similar small effect. Overall, including correlations in the averaging procedure for the 3NF should bring the results closer to the empirical saturation point. 

\begin{figure}
\begin{center}
\includegraphics[scale=0.55]{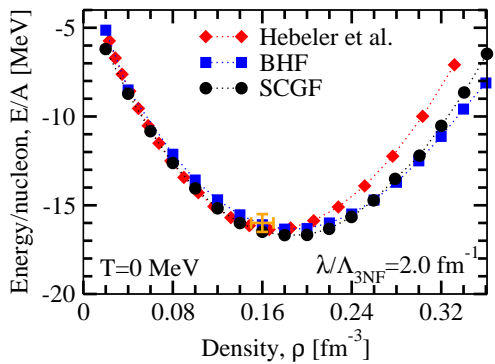}
\caption{(Color online) Comparison of results for the energy per nucleon of SNM obtained with different approaches using the same SRG-evolved 2NF and a 3NF. Circles correspond to extrapolated SCGF results, wheres squares are BHF calculations at $T=0$ MeV. Diamonds correspond to the results of Hebeler \emph{et al.} \cite{Heb11}}
\label{fig:ener_comp_hebeler}
\end{center}
\end{figure}

Second, and as already been pointed out, calculations with the SCGF approach are performed at finite temperature to avoid pairing instabilities. Thermal effects can be estimated using a variety of ways. BHF zero-temperature calculations show that thermal correlations are somewhat repulsive and of the order $\sim 1$ MeV at saturation, in accordance with the Sommerfeld expansion. Even at the lowest density considered here, where temperature should have the largest effect, the $T=0$ and $T=5$ MeV energies differ by only $\sim 3$ MeV. We will present in Fig.~\ref{fig:ener_comp_hebeler} results which have been extrapolated to zero temperature by means of a simple rule. Thermal effects are thus under control and allow us to explore the finite-temperature phase diagram of  nuclear matter. 

\begin{figure}
\begin{center}
\includegraphics[scale=0.55]{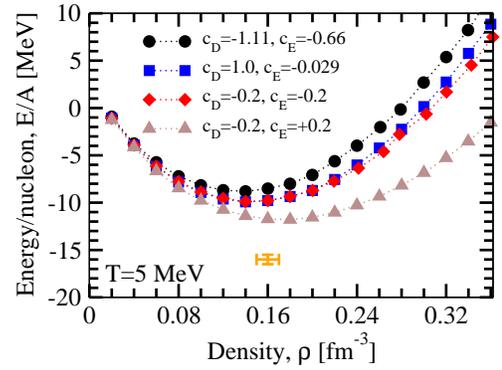}
\caption{(Color online) SCGF results for the energy per nucleon of SNM as a function of the density at a temperature of $T=5$ MeV for different LECs. The different combinations are discussed in the text. }
\label{fig:ener_cdce}
\end{center}
\end{figure}

Third, within a given renormalization scale, one might expect quantitative changes to the saturation properties coming from different determinations of the LECs. We explore these differences in Fig.~\ref{fig:ener_cdce}, where we show the saturation curve at $T=5$ MeV obtained with four different combinations of $c_D$ and $c_E$. In all cases, the unevolved Entem-Machleidt N3LO potential has been complemented with a density-dependent force with the same functional dependence but different LECs. The density-dependent 2NF is computed in all cases with $\Lambda_\chi=700$ MeV and a regulator with $\Lambda_\textrm{low k}=500$ MeV. 

The choice of combinations is representative of the spread in LEC values associated to different fitting protocols. The first choice, $c_D=-1.11$ and $c_E=-0.66$ (circles), has been determined from the binding energies of $^3$H and $^4$He in exact few-body calculations \cite{Nogga2006}. The second set, $c_D=1.0$ and $c_E=-0.029$, (squares) has been obtained with a local version of the 3NF, as often used in finite nucleus calculations \cite{Cip13}, but fit to the $A=3$ system only \cite{Nav07}. The third set of LECs, with $c_D=-0.2$ and $c_E=+0.2$ (triangles), has been used in Ref.~\cite{Krewald2012} for SNM calculations on the basis of naturalness and provide further typical values of these constants. 

Curves corresponding to different values of $c_D$ and $c_E$ fall within a narrow band below saturation. In general, the changes are mild even if the constants are changed considerably, indicating that 3NF contributions are small at low densities. Above $\rho \sim 0.16$ fm$^{-3}$, however, differences develop as density increases. 
These differences can be largely explained by the contact term and its contribution to the total energy, proportional to $c_E$. In the Hartree-Fock approximation, this reads:
\beq
\frac{E_{c_E}}{A} = - \frac{3}{16} \frac{c_E}{f_\pi^4 \Lambda_\chi} \rho^2 = -5.5 \, c_E \left( \frac{\rho}{\rho_0} \right)^2 \text{ MeV.}
\enq
One  therefore expects negative values of $c_E$ to lead to more repulsive contributions. To this purpose we use in Fig.~\ref{fig:ener_cdce} a fourth set of LECs,  $c_D=-0.2$ and $c_E=-0.2$ (diamonds), which differs by a minus sign in $c_E$ with respect to the third set. As expected, positive values for $c_E$ yield results closer to the saturation point, but also have smaller compressibilities.
Note that, in general, the saturation point is only marginally dependent on the choice of constants for $c_E<0$. 
In all cases, saturation lies far from the empirical value. Further improvements in the treatment of the 3NF are expected to damp the repulsive effect of 3NF and thus bring the results closer to the empirical saturation point. Alternatively, one could try to obtain new values of the LECs by fitting the saturation point itself \cite{Krewald2012}. Note that higher orders in perturbation theory for the 3NF can also modify this picture \cite{Kaiser2012}.

A fourth source of theoretical error in our calculations is the order at which $\chi$EFT is implemented in the interaction. A consistent measure of errors would only be provided if results where compared order by order in perturbation theory. In other words, NLO, N2LO and N3LO results should yield smaller and smaller error bands. For the results presented up to now, however, we have considered different order for each force, as the averaged 3NF comes from an N2LO interaction. At N3LO, one would in principle also have to include four nucleon forces, as recently done for neutron matter in Ref.~\cite{Tews13,Krug13}. 

\begin{figure}
\begin{center}
\includegraphics[scale=0.55]{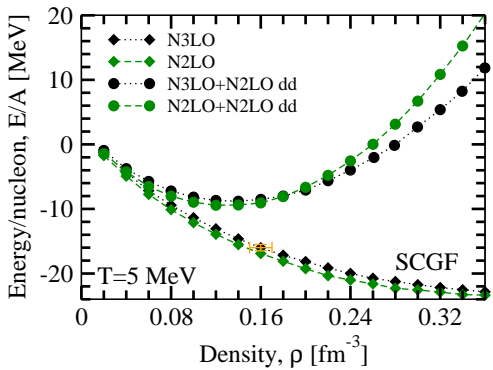}
\caption{(Color online) SCGF results for the energy per nucleon of SNM as a function of the density at a temperature of $T=5$ MeV, using N3LO (dotted lines) and the optimized version of N2LO (dashed lines) from \cite{Eks13}.  Diamonds represent calculations including only 2NF, whereas circles include 2NF and 3NF as described in the text.}
\label{fig:ener_nnloopt}
\end{center}
\end{figure}

Lately, an optimized version of the 2NF at N2LO has been obtained by optimizing the LECs to scattering data using a derivative-free minimization procedure \cite{Eks13}. The use of this force enables us to be fully consistent from the point of view of the order of perturbation in the chiral expansion. In Fig.~\ref{fig:ener_nnloopt}, we compare the energy per particle at $T=5$ MeV using the two-body Entem-Machleidt potential (dashed lines) \cite{Ent03} and the optimized two-body N2LO of Ref.~\cite{Eks13} (dotted lines). Results considering both 2NF (diamonds) and averaged 3NF (circles) are showed. For the N2LO 3NF, we use the values of $c_D$ and $c_E$ quoted in Ref.~\cite{Eks13}. At the level of two-body interactions, the results are very similar, with N2LO consistently more attractive than N3LO in the whole density range. This behavior is in fair agreement with the results presented in Ref.~\cite{Baar2013}.

As already observed in Fig.~\ref{ener}, the inclusion of 3NF is crucial to obtain saturation at realistic densities and energies. The new optimized force yields slightly more attractive results at low densities, but becomes more repulsive above saturation. This indicates a larger compressibility for the N2LO force. In addition, the saturation density with the optimized N2LO is smaller, but the saturation energy is slightly more attractive with respect to N3LO. This behavior follows the same trend observed for calculations in neutron matter \cite{Eks13,Baar2013}. Overall, it can be ascribed to the different structure of the 2NF itself. Note that the N2LO potential yields a poorer reproduction of the phase shifts for selected partial waves compared to the richer N3LO force.

Most nuclear matter calculations using chiral forces have been performed in a perturbative framework starting from evolved interactions. In Ref.~\cite{Heb11}, convergence has been analyzed order by order in many-body perturbation theory. Results have been obtained up to third order, including particle-particle and hole-hole propagation \cite{Heb11}. In principle, the equation of state should be independent of the evolution scales in the 2NF and the 3NF. Moreover, in the perturbative regime, results should only be mildly dependent on the order in perturbation theory. Our non-perturbative calculations include contributions to all orders and hence are neither limited to the perturbative regime nor dependent on the order of perturbation theory. If the diagrammatic summation is complete, it should lead to scale-invariant results.

\begin{figure}
\begin{center}
\includegraphics[scale=0.55]{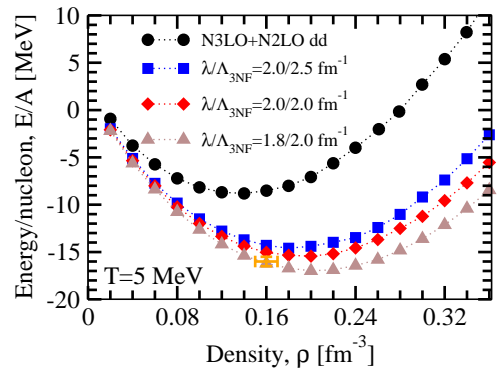}
\caption{(Color online) 
SCGF results for the energy per nucleon of SNM as a function of the density at a temperature of $T=5$ MeV. Different lines represent different choices of cut-offs for the 2NF, $\lambda$, and the 3NF, $\Lambda_\text{3NF}$. }
\label{ener_srg}
\end{center}
\end{figure}

We test this hypothesis by performing calculations at different evolution scales, both in the two- and the three-body sector. We evolve the 2NF using a free space SRG transformation \cite{Bog07}. 
The transformation renormalizes the 2NF, suppressing off-diagonal matrix elements and giving rise to a universal low-momentum interaction. The SRG evolution flow also induces many-body forces, which should be taken into account to keep the calculation complete. Following the philosophy of Ref.~\cite{Heb11}, we incorporate the effect of induced forces through the re-fitting of the $c_D$ and $c_E$ LECs to the $^3$H binding energy and $^4$He matter radius. We use the values given in Table I of \cite{Heb11}. Note that this assumes that the operatorial and momentum structures of the original and the induced 3NFs are the same. Furthermore, we explore the dependence of our results on the 3NF cut-off, $\Lambda_\mathrm{3NF}$, appearing in the density-dependent 2NF. A more complete calculation would require running an SRG evolution including the 3NF \cite{Heb13,*Wendt2013}. 

We present the results of this exploration in Fig.~\ref{ener_srg}. Numerical calculations obtained using the SRG on the 2NF have a saturation point which is quite closer to the empirical value when compared to the original force. Moreover, if the 2NF has been SRG-evolved, the results are somewhat independent of the cut-off. Overall, one can say that the more the 2NF is evolved down, the more attractive the saturation curve becomes. This effect is a consequence of the shift in importance between the 2NF and the induced 3NF associated to the SRG. There is also a small dependence on $\Lambda_\text{3NF}$, but the differences agree well with those presented in Ref.~\cite{Heb11}.

The large differences between the results obtained with evolved and unevolved forces is striking. If correlations and induced many-body forces had been fully taken into account, one would have expected a much closer agreement between the results. This difference might indicate that the assumptions associated to induced 3NF are not necessarily robust. Missing induced three-body forces, which up to now have not been included in SNM calculations, could solve this discrepancy. Alternatively, the difference is also an indication of missing many-body effects as, for instance, higher orders in the treatment of the 3NF. 
It must be emphasized that the present way to proceed when applying SRG evolution in infinite matter should be improved, carrying out the evolution on a full hamiltonian with both two- and three-body forces. Recently, improvements towards the solution of this problem have been presented for calculations in pure neutron matter \cite{Heb13}, where a full hamiltonian has been consistently evolved.
All in all, our results seem to contradict the idea that induced 3NFs can be treated simply in nuclear matter.

In terms of evolved interactions, our non-perturbative calculations can be used to check whether the perturbative regime is actually reached. To this end, we compare, in Fig.~\ref{fig:ener_comp_hebeler}, our results to the perturbative calculations presented in Ref.~\cite{Heb11}. The BHF and SCGF calculations have been performed with an SRG-evolved 2NF and a 3NF with the same cut-offs, $\lm/\Lambda_\mathrm{3NF}=2.0/2.0$ fm$^{-1}$. Whereas the Brueckner results have been obtained with a zero temperature code, the SCGF calculations have been extrapolated to zero temperature by means of a simple procedure. At low temperatures, the Sommerfeld expansion indicates that the effect of temperature is quadratic and the same, but with opposite sign, for the energy and the free energy \cite{Rios2008}.  Consequently, the semi-sum of both thermodynamical potentials is an estimate of the zero-temperature energy. 
We obtain an extremely good agreement between both many-body approaches and the perturbative calculations of Ref.~\cite{Heb11} in a wide density range. Up to saturation, the results fall in a narrow band of $\sim 0.5-0.1$ MeV, indicating that the perturbative regime is indeed achieved \cite{Gez13}. Above $\rho \sim 0.20$ fm$^{-3}$, perturbation theory results become more repulsive than the SCGF and BHF predictions.  Note that the difference is still rather small, certainly smaller than that usually obtained in comparisons of different many-body methods and 2NF \cite{Baldo2012}. A possible source for the discrepancy could be the somewhat different treatment of the density-dependent 2NF  in Refs.~\cite{Heb11} and \cite{Holt10}. Alternatively, the difference might indicate that the effect of higher (beyond third) orders in perturbation theory becomes non-negligible at densities beyond saturation. Both the SCGF and the BHF results agree beyond this point, demonstrating that non-perturbative effects are well described within a ladder-type resummation.

\subsection{Single-particle properties}

\begin{figure}[t!]
\begin{center}
\includegraphics[scale=0.55]{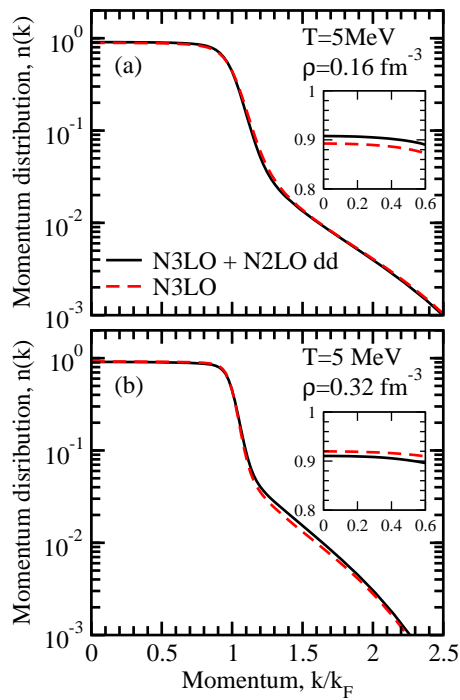}
\caption{(Color online) Momentum distribution in SNM using  2NF only at N3LO (dashed line) and using the density-dependent 2NF (solid line). Calculations are performed at the empirical saturation point (upper panel, (a)) and at twice this density (lower panel, (b)). The insets focus on the depletion below the Fermi surface.}
\label{fig:mom_dist}
\end{center}
\end{figure}

SCGF calculations give direct access to the microscopic properties related to the SP propagator. These include self-energies, spectral functions and momentum distributions, from which one can derive microscopic, transport and bulk properties \cite{Rios2008}. In particular, the SP momentum distribution of Eq.~(\ref{mom}) provides a measure of the correlations embedded in the nuclear wave function.
In Fig.~\ref{fig:mom_dist}, we show $n(k)$ at  $T=5$ MeV for two densities, $\rho$=0.16 fm$^{-3}$ (upper panel) and
$\rho$=0.32 fm$^{-3}$ (lower panel). We compare results obtained with the 2NF alone (dashed line) and with the density-dependent 2NF (solid line).
The effect of the 3NF in the momentum distribution is relatively small around saturation. We need to go to higher densities to appreciate a noticeable difference in $n(k)$. In fact, in spite of the fact that 3NFs have a large repulsive effect in the energy as density increases, the effect on the momentum distribution is rather mild. As demonstrated in Fig.~\ref{fig:mom_dist}, high momentum components hardly change at all in a region up to $0.32$ fm$^{-3}$.

3NFs induce a somewhat stronger density dependence in the depletion of the momentum distribution. This is shown in the insets of the Figure, which focus on the low-momentum region. The somewhat soft chiral 2NF induces a relatively small depletion, of order $10 \%$, as compared to traditional 2NF, which typically have $~15-20 \%$. For these traditional two-body potentials, the density dependence of the depletion is generally very soft and dominated by tensor correlations \cite{Rios2009a}. 3NF modify this behavior, possibly due to the additional tensor structures associated to the density-dependent 2NF \cite{Holt10}. At sub-saturation densities, the 3NF decreases the depletion. The difference is small at $\rho$=0.16 fm$^{-3}$ (upper panel in Fig.~\ref{fig:mom_dist}), within a few percent. We observe a higher depletion when considering only 2NF, \emph{i.e.} $n(0)_\mathrm{2NF}=0.892$ versus $n(0)_\mathrm{2NF+2NFdd}=0.908$.  This causes a slightly higher kinetic energy for the system without 3NF, that is, $T_\mathrm{2NF}=32.88$ MeV to be compared with $T_\mathrm{2NF+2NFdd}=32.23$ MeV. At higher densities (see $\rho$=0.32 fm$^{-3}$ in the lower panel), the 3NF induce a slightly larger depletion,  $n(0)_\mathrm{2NF+2NFdd}=0.910$, compared with $n(0)_\mathrm{2NF}=0.919$. Correspondingly the kinetic energy in the system with 3NF, $T_\mathrm{2NF+2NFdd}=46.76$ MeV, is a little bit higher than that with 2NF only, $T_\mathrm{2NF}=45.44$ MeV.

\begin{figure}
\begin{center}
\includegraphics[scale=0.46]{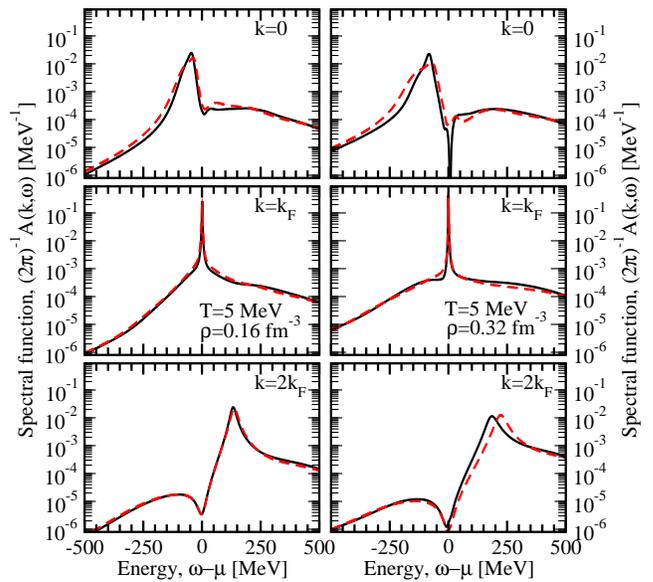}
\caption{(Color online) SP spectral function for SNM using only 2NF at N3LO (dashed line) and including the density-dependent 2NF (solid line). Calculations are performed at the empirical saturation density (left panels) and at twice this density (right panels). Results are displayed for three characteristic SP momenta. }
\label{fig:spect}
\end{center}
\end{figure}

One can understand the small changes in the momentum distribution by looking at the effects of the density-dependent 2NF on the SP spectral function. Fig.~\ref{fig:spect} shows the spectral function for three characteristic momenta, $k=0$, $k=k_\mathrm F$ (Fermi momentum) and $2k_\mathrm F$. The left (right) panel corresponds to a density of $\rho=0.16$ fm$^{-3}$ ($\rho=0.32$ fm$^{-3}$). The effect of the 3NF force on the spectral function is relatively small. In general, the strength beyond the quasi-particle is hardly affected by 3NFs. This suggests that 3NF effects can be well described, within a quasi-particle picture, by shifts in the quasi-particle spectra. More specifically, at low momenta, 3NFs tend to narrow the quasi-particle peak and increase its height, but do not change its energy. Around the Fermi surface, the difference is hardly noticeable.

The largest effect appears above the Fermi surface, where the quasi-particle peak is shifted to lower energies. Naively, one might expect an attractive effect on the saturation curve when including the 3NF. Note, however, that the energy dependence is plotted with respect to the chemical potential, $\mu$. The variation in the latter is larger than the quasi-particle shift and results in an overall repulsive effect of the 3NF as the density increases. Our results agree qualitatively with previous SCGF studies, in spite of the different interactions and averaging procedures \cite{Soma08,Soma09,*SomaPhD}. Note that these previous calculations did not include the different pre-factors in the self-energy, but had an energy estimate obtained from the $T$-matrix itself rather than the GMK sum-rule.

At this point, let us stress the fact that correlations have a substantial effect in the SP properties at all momenta. In spite of the cut-off in both the 2NF and 3NF considered here, one finds a substantial population of high momentum components. Similarly,  the spectral functions display qualitatively important tails at high energies. Traditional microscopic 2NF would yield even larger high-momentum components and fragmentation. The interplay between tensor and short-range correlations is at the basis of the renormalization of strength. Our calculations indicate the importance of considering such effects in many-body calculations even with relatively soft interactions. In particular, let us stress that low-momentum SP properties are affected by correlations. This is a direct effect of self-consistency and provides a feedback mechanism, whereby high-momentum modes affect low energy properties. 

\begin{figure}
\begin{center}
\includegraphics[scale=0.55]{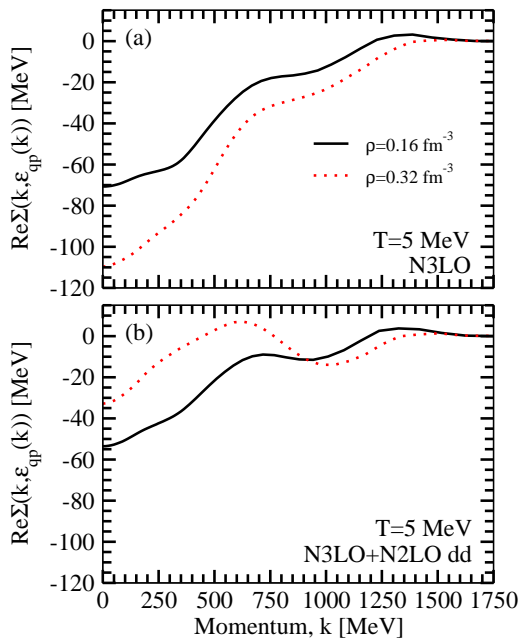}
\caption{(Color online) SP potential for SNM using only 2NF at N3LO (upper panel, (a)) and including the density-dependent 2NF (lower panel, (b)). Calculations are performed at the empirical saturation density (solid line) and at twice this density (dotted line). }
\label{fig:sppot}
\end{center}
\end{figure}

3NF have a particularly strong effect on the quasi-particle spectra. The latter is obtained in our approach by solving the implicit equation:
\beq
\epsilon_\text{qp}(k) = \frac{k^2}{2m} + \text{Re}\Sigma(k,\epsilon_\text{qp}(k)) \, .
\enq
We show in Fig.~\ref{fig:sppot} the potential part of the quasi-particle energy, \emph{i.e.} the second term in the previous equation. This has been calculated at saturation density and twice its value with both 2NF (upper panel) and 3NF (lower panel). At saturation density, $\rho=0.16$ fm$^{-3}$, 3NF shifts the quasi-particle strength to more repulsive values. At zero momentum, the shift amounts to $\sim 15$ MeV, but it gradually disappears as momentum increases. Overall, this indicates that 3NF effects are more important in the low-momentum regime. At $\rho=0.32$ fm$^{-3}$, 3NF also affect the qualitative momentum dependence of the SP potential, which even becomes repulsive at intermediate momenta. Overall, we observe a much stronger repulsion at low momenta, of $\sim 75$ MeV at zero momentum. This strong density-dependent repulsive effect drives the saturation of the total energy of the system, as observed in Fig.~\ref{ener}.

The effect of the cut-off in the underlying interactions is clearly seen in the large momentum behavior of the SP potential. Calculations with traditional interactions show a steady increase of the potential with momentum, whereas chiral interactions indicate a damping of the SP potential at momenta above $\sim 1500$ MeV. In a Hartree-Fock picture, a sharp cut-off, $\Lambda$, in relative momentum matrix elements will affect the self-energy in a SP momentum  region starting from  $2 \Lambda - k_F$ on. Specifically, above $2 \Lambda + k_F \sim 1250-1500$ MeV, the self-energy is expected to be zero in this simple approach. This indicates that the damping of the SP potential energy as well as the structures in the momentum region $~\sim 700-1500$ MeV are due to cut-off in the original interaction.

\section{Conclusions}

We have performed nuclear many-body calculations for infinite nuclear matter using 2NF and 3NF arising from $\chi$EFT. To this end, we have extended the SCFG theory within the ladder approximation to include 3NF via a density-dependent 2NF. We use one- and two-body effective interactions, that generalize the concept of normal ordering to correlated ground states. For this initial exploratory study, we have used a density-dependent 2NF from an uncorrelated average over the third particle, as described in Ref.~\cite{Holt10}. We have implemented an equivalent method within the BHF approach, taking into account corrections in the SP self-energy and in the calculation of the total energy by means of the Koltun sum-rule.

Saturation properties of SNM improve substantially when considering a density-dependent 2NF in the many-body calculation. This is true for both the SCGF and the BHF methods. In chiral interactions, LECs in the two-nucleon sector have been fit to scattering and deuteron data. The remaining two LECs have to be fit to the few-body sector. We have explored the dependence of the saturation curve to different determinations of LEC. The change in saturation properties with $c_D$ and $c_E$ in a range of natural values  is relatively mild. The differences can be largely explained in terms of the relative importance of the contact term. 

In order to be fully consistent in the order of perturbation in the chiral expansion, we have also performed calculations using the newly optimized N2LO force, which induces  a slight repulsion in the saturation curve of nuclear matter with respect to the N3LO results. This agrees with recent results obtained using the coupled cluster approach. 

Moreover, our results have been compared to cases when an SRG evolution is applied to the 2NF. Our motivation has been to test if induced 3NF can be effectively included as a renormalization of LECs. Our results show a large difference between the saturation curve obtained with original, unevolved, forces and SRG-evolved results. These differences might arise from the assumption that induced three-body forces can be mocked by a refitting of the LECs. This would indicate that induced 3NF play an important role in infinite matter, similarly to what is observed in finite nuclei \cite{Cip13}. In terms of evolved hamiltonians, we find similar results in the infinite system independently of the cut-off after the LECs have been re-fitted. We have also found a very good agreement between our many-body calculations and the perturbative results of Ref.~\cite{Heb11}.

The SCGF also provides the single-particle, microscopic properties of the system. We have discussed here the momentum distribution, which is an indicator of correlations. Chiral interactions consistently provide a low-momentum  depletion of $\sim 10 \, \%$. Our results suggest that the momentum distribution is hardly affected by 3NF in the high-momentum region. At low momenta, 3NF seem to induce a larger density dependence on the depletion below the Fermi surface. We have also discussed the sensitivity of the spectral function to 3NF. We have found a rather small dependence when the spectral function is plotted as a function of the energy shifted with respect to the chemical potential. The widths and heights of the quasi-particle peaks are hardly modified, but we observe a shift in the quasi-particle peak at high momenta. To better appreciate the effect of 3NF at a microscopic level we looked at the SP potential. The latter is affected by 3NF mainly at low momenta, where it induces a strong repulsion, especially at high densities.

We plan to develop our calculations with a series of improvements in the near future. To start with, we want to improve upon the one-body average over the third particle. We expect that correlations shall play a role in some of the terms in the density-dependent 2NF. Specifically, we want to include off-diagonal matrix elements in relative momenta as well as a dressed one-body propagator in the average. The latter effect plays an important role in finite nuclei \cite{Cip13}. Further, and for completeness, we would like to account for the two-body average associated to the four-point propagator. 

In terms of physical applications, we also expect the equation of state of pure neutron matter to be substantially modified by 3NF. Our non-perturbative scheme would allow us to extend the density region of applicability of many-body theory using chiral interactions \cite{Heb11}. This could lead to a reduction of the uncertainty band in the mass-radius relation \cite{Hebeler2012}. Our calculations can also provide superfluid and transport properties of neutron-star matter. More importantly, we can also obtain microscopic single-particle information and hence describe the evolution of correlations with isospin asymmetry in dense systems. Ultimately, we aim at a complete \emph{ab initio} description of nuclear systems starting from chiral interactions.

\acknowledgments

We thank K. Hebeler for providing us with the data of Fig.~\ref{fig:ener_comp_hebeler}.
We also thank C. Barbieri and J. W. Holt for enlightening discussions. 
This work is supported by the Consolider Ingenio 2010 Programme CPAN CSD2007-00042, 
Grant No.~FIS2008-01661 from MEC and FEDER (Spain) and 
Grant No.~2009GR-1289 from Generalitat de Catalunya  (Spain)
and by STFC, through Grants ST/I005528/1 and ST/J000051/1. 

\bibliographystyle{apsrev4-1}
\bibliography{biblio}

\end{document}